# Recommender Systems in E-commerce

Tanmayee Salunke
Department of Computing and Informatics
Bournemouth University
*Bournemouth, United Kingdom*
s5431759@bournemouth.ac.uk

Unnati Nichite
Department of Computing and Informatics
Bournemouth University
*Bournemouth, United Kingdom*
s5439414@bournemouth.ac.uk

*Abstract*— E-commerce recommender systems are becoming increasingly important in the current digital world. They are used to personalize user experience, help customers find what they need quickly and efficiently, and increase revenue for the business. However, there are several challenges associated with big data-based e-commerce recommender systems. These challenges include limited resources, data validity period, cold start, long tail problem, scalability. In this paper, we discuss the challenges and potential solutions to overcome these challenges. We also discuss the different types of e-commerce recommender systems, their advantages, and disadvantages. We conclude with some future research directions to improve the performance of e-commerce recommender systems.

## I. INTRODUCTION

Whinston and Kalakota describe e-commerce as "the process of acquiring and marketing information, goods and services using computer networks, which are usually the Internet[1]. Ecommerce businesses employ recommender systems to direct customers toward goods that are most likely to be purchased by them. They enable e-commerce site customisation, which aids in generating sales, and they automate personalization on the Web. These suggestions are determined using the customer's demographics, the manner and content of previous purchases, and the top-rated merchants on the website[2].

Both customised and non-personalized recommender systems are available. The former is based on a user's choices (favourite book or music genre) and is one of the more effective ways to produce suggestions and services. These are frequently contrasted with what industry experts believe will best serve the client given their preferences. Basically, the personalised recommender system may produce recommendations based on the varied interests of the users, which not only cuts down on the amount of time spent searching but also helps e-commerce companies increase their sales [3].

The primary goal of this study is to investigate different types of recommender systems their advantages, disadvantages, challenges faced by the recommender systems, how to overcome the challenges, use case study. Industry giants utilise this technology to assist clients in finding the goods or services they need and in making the best choice.

## II. LITERATURE SURVEY

A user may find it challenging to locate a product that meets his requirements and is priced reasonably when the number of e-commerce websites rises due to the large number of products available. Hence, the consumer has no other choice but to search for each product on each website until he finds the one, he needs in order to locate the greatest product. The author of this study [5] proposed a system that makes it simple for the user to choose the product he wants while saving him time by allowing him to see the results in real-time. More particular, the lengthy process in which customers frequently find themselves caught is catered for by this system approach. Considering time is so valuable, this system will be focused on giving the user the solutions to their queries quickly and effectively[5]. In order to provide recommendations for user-generated contents, Z. Wang et al.[6] suggested combining social online networks and information sharing networks. The author recommended the videos that the user is most likely to actually or buy on the social networking website. An updated user-content matrix technique is suggested to forecast how the videos might be shared or imported. Utilizing social interactions, user activity, similarity of content the author created a shared user-content space based on this methodology. These experiment results, which are verified by Weibo traces show the benefit of fusing suggestions based on social and content. Comparing this to the current content-based filtering and collaborative filtering technique, suggestion accuracy is significantly higher. According to Brent Smith et al.[7], Amazon.com developed its collaborative item-based filtering and recommendation system in 1998, which ignores user history. It overcomes difficulties presented by new clients or infrequent clients who little or no history. Based on the user's prior behaviour and present situation, this tip makes specific recommendations for limited goods. Despite spending money on non-media things, it increased Amazon's sales by following this trend. In this paper [8], To increase the accuracy of recommendations and take into account how social networks affect them, the author suggested a method based on ratings, reviews, and social relationships. The LR model is used to forecast ratings, CoDA is used to find communities, and Using Word2Vector to convert text revision into vectors [8]. This experiment compares a traditional model of matrix factorization, neighbourhood models, and a cutting-edge using a social network, enhancing the model's accuracy.

Takuma et al. [9] developed a manual base library of words in review texts using the "MeCab" analysis engine quite frequently, which has an automatic noun extraction feature that are present in evaluation. By comparing the user's characteristics to those of the donors, a user might be recommended a list of hotels. The most significant problems with RS were outlined by Mohamed et al. [10] in a very good survey that provided examples of alternatives from the most recent research at the time. They did not, however, address the evaluation of RS as a difficulty or how the features of the data can impact RS performance.

## III. USe Case study of Recommender Systems based on E-commerce

### A. Actors:

*1) Customer:* A customer of an E-commerce website who wants to find new products.

*2) Recommender System:* The system that provides recommendations for customers based on their purchase history and preferences.

### B. Preconditions:
• Customer has browsed or purchased items from the E-commerce website.
• The customer has a profile with the website that contains their purchase history and preferences.

### C. Basic Flow of Events:

*1)* Customer visits the E-commerce website.

*2)* The website recognizes the customer based on their profile.

*3)* The recommender system analyzes the customer's purchase history and preferences to generate a list of recommendations.

*4)* The recommendations are presented to the customer on the website.

*5)* The customer can browse the recommendations and choose which items they want to purchase.

### D. Postconditions:

*1)* The customer has purchased the items they wanted.

*2)* The website's recommendation engine has been updated with the customer's new purchase history and preferences.

## IV. Types of Recommender System

### A. Collaborative Filtering: -

A collaborative recommender system is a recommender system that relies on the ability of a group of users to pool their knowledge and make recommendations. It is founded on the notion that a community's collective intellect of users is more powerful than any single individual. The collective intelligence of a community can be harnessed to make better recommendations than any single individual. The collaborative recommender system relies on the community's combined wisdom to generate suggestions. The community's collective intellect is more powerful than any single individual. The community can harness the collective intelligence to make better recommendations than any single individual[4].

*1) Working of Collaborative Filtering:*

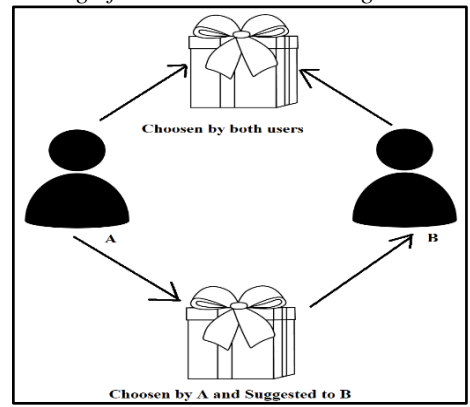

Collaborative recommender systems are those that make suggestions made to a user based on other users' preferences. The assumption is that if two users have similar preferences, then they are more likely to like the same items. These systems typically use some form of collaborative filtering, which is a process of predicting an individual's interests based on the interests of other users. User-based and item-based collaborative filtering are the two primary varieties. In user-based collaborative filtering, suggestions are presented to a user based on their shared preferences. That is, if User A and User B have similar preferences, then the system will recommend items to User A that User B has liked. Item-based collaborative filtering is where recommendations are made to a user based on the items that the user has liked in the past. That is, if User A has liked item 1 and item 2, and User B has also liked item 1 and item 2, then the system will recommend to User A items that User B has liked.

*2) Advantages:*

*a)* It can provide more accurate recommendations than a single recommender system. This is because it can learn from the feedback of multiple users and identify patterns that a single recommender system may not be able to recognise.

*b)* It can provide recommendations for a wide range of items. This is because it can learn from the feedback of multiple users and identify patterns that a single recommender system may not be able to identify.

*c)* It can offer suggestions for a wide range of users. This is because it can learn from the feedback of multiple users and identify patterns that a single recommender system might not be able to recognise.

*3) Disadvantages:*

*a) Users are reluctant to disclose their information:* Users are reluctant to share their information because of privacy issues.

*b) Users tend to be myopic:* Users tend to focus on the short-term and do not consider long-term benefits.

*c) Social loafing:* Social loafing is a problem that often occurs in collaborative recommender systems. It refers to the phenomenon that users are less likely to contribute when they know that others are also contributing.

B. *Content based: -*

Content filtering is a technique used to suggest products to customers of a system according to the similarity of the items. It is commonly used in recommender systems to find similar items to those that a user has already expressed interest in. The content that is used to determine the similarity of items can be anything from text to images to ratings. The goal of a user's with a recommender system based on content screening recommendations for products that are similar to those that they have already expressed interest in.

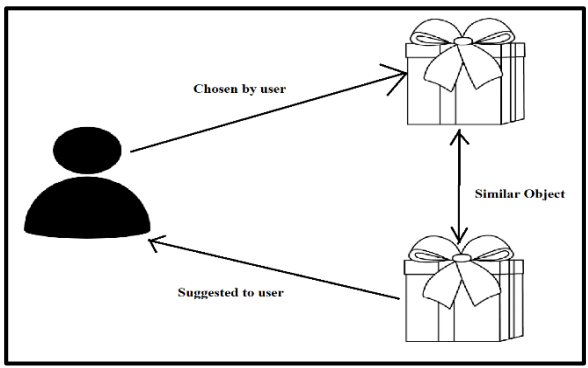

A content-based e-commerce the item recommendations made by the recommender system are similarity between the ratings. That is, the system recommends products that are comparable to those the consumer has already purchased. The system first calculates the similarity between the things using the ratings of the things. The ratings of the items can be the plot. The system then recommends the things that are most similar to the things that the user has purchased. Content-based recommender systems are easy to build and they do not require a lot of data. However, they are limited in that they do not take into account the Individual preferences. That is, the system does not know if the user likes costly, average or low-price things.

*1) Advantages:*

*a)* Increased accuracy: Content-based filtering algorithms are ability to provide users product recommendations with a high degree of accuracy, as they are able to learn the user's preferences from the past behaviors and use this information to make recommendations.

*b)* Improved relevance: Content-based filtering algorithms are able to suggest products that are more relevant to the user, as they are able to take into account the user's past behaviors when making recommendations.

*c)* Increased serendipity: Content-based filtering algorithms often recommend items that the user would not have thought to search for, which can lead to increased serendipity and a broader range of items being recommended.

*d)* Reduced noise: Content-based filtering algorithms are able to reduce the amount of noise in the recommendations, as they are able to filter out items that are not relevant to the user.

*2) Disadvantages:*

*a)* It can be difficult to determine the similarity of items.

*b)* It can be time-consuming to compute recommendations.

*c)* The quality of recommendations may be sensitive to the choice of similarity metric.

*d)* The quality of recommendations may be sensitive to the choice of filtering method.

*e)* It can be difficult to scale content filtering to a large number of items.

C. *Demographics Based:*

A recommender system based on demographics would take into account factors such as age, gender, location, and interests when making recommendations. It would use this information to recommend items that are popular with people who have similar demographics.

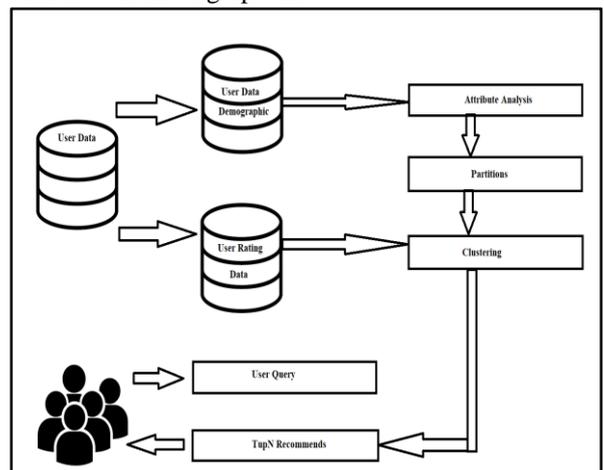

In a e-commerce recommender system based on demographics, the system would take in demographic information about users in order to make better recommendations. This could include information like age, gender, location, and interests. The system would then use this knowledge to suggest products that are popular with people who share similar demographics. For example, if the system knows that a user is a 25-year-old female from the United States, it would recommend item that are popular with other 25-year-old females from the United States. This would ensure that the recommendations are tailored to the user's specific interests. Overall, using demographics in a recommender for e-commerce system can be a helpful way to provide more accurate advice. However, it is important to consider both the advantages and disadvantages before implementing such a system.

*1) Advanatages:*

*a)* It can provide more accurate recommendations to users. This is because the system can take into account factors such as age, gender, and interests when making recommendations.

*b)* This type of system can be used to target ads and content to specific demographics, which can be beneficial for businesses.

*2) Disadvantages:*

*a)* It is possible that the demographics of the users of the system do not accurately reflect the diversity of opinions and preferences in the population at large. This could lead to the system recommending items that are not of interest to a large number of people.

*b)* Additionally, if the system relies too heavily on demographics, it could fail to recommend items that are popular with people outside of the target demographic, leading to a less diverse and interesting set of recommendations.

*c)* Finally, if the system is not well-designed, it could result in recommendations that are biased in favor of certain groups of people or that exclude certain groups altogether.

*D. Community Based:*

A community-based recommender system is a kind of recommender system that counts on the user base for suggestion input. This type of system is often used in social networking sites, where users are connected to each other and can provide recommendations to each other.

A community-based item recommender system is a system that relies on the collective feedback of a community of users to provide recommendations. The system first collects ratings and reviews from a group of users, then uses this data to generate recommendations for solitary users. The recommendations are founded on collective preferences of the community, rather than the individual preferences of any one user[11].

A community-based recommender system can also provide a more diverse set of recommendations than a system that relies on individual user preferences. Because the system is not limited to the preferences of any one user, it can recommend a wider range of things. This can be especially helpful for users who have eclectic taste in shopping, or who are looking for something outside of their usual choice.

Overall, a community-based recommender system has the potential to be more accurate and diverse than a system that relies on individual user preferences. However, the system is only as good as the data that is input into it, so it requires a large and active community of users to be effective.

*1) Advantages:*

*a)* It helps to provide personalized recommendations to the users.

*b)* It helps to raise the standards for suggestions.

*c)* It helps to make the data's dimensions smaller.

*d)* It aids in enhancing the accuracy a list of suggestions.

*2) Disadvantage:*

*a)* There is the potential for a negative feedback loop, where users only see recommendations for items that are popular within the community, regardless of whether or not they would actually enjoy those items.

*b)* There is also the potential for echo chambers, where users only see recommendations for items that confirm their existing biases and preferences.

*c)* Community-based recommender systems can also be susceptible to gaming, where users artificially inflate the ratings or popularity of certain items in order to manipulate the recommendations.

*E. Hybrid Recommender System*

This hybrid a mix of collaborative filtering and content-based filtering makes up the e-commerce recommender system.. Content-based filtering uses the product information such as product category, brand, and product description to recommend similar products to the user. Collaborative filtering uses the user's purchase history and reviews from other customers to recommend products to the user.

The system first collects data from the e-commerce site, such as product information, customer reviews, and customer purchase history. This data is then used to create a content-based model using machine learning methods and natural language processing. This model will be used to generate product recommendations based on the user's search query or product description.

The system then uses the customer reviews and purchase history to create a collaborative filtering model. This model will generate product recommendations based on the user's past purchases, reviews, and ratings.

The hybrid e-commerce recommender system will then combine the content-based and collaborative filtering models to generate the most relevant product recommendations for the user. The system will use the customer purchase history and reviews to refine the content-based recommendations. This will ensure that the user is presented with the most appropriate recommendations.

*1) Advantages: -*

*a) Increased Customer Engagement:* Hybrid e-commerce recommendation systems use a combination of both collaborative and content-based filtering techniques to generate more accurate recommendations. This means that customers are presented with a variety of product recommendations customised to meet their particular demands and interests. This increases the chances of customers engaging with the website and making a purchase.

*b) Increased Revenues:* Since customers are presented with more personalised product recommendations, there is an increased likelihood that they will purchase the recommended items. This leads to increased revenues for the company.

*c) Improved User Experience:* Hybrid e-commerce recommendation systems use a combination of both collaborative and content-based filtering techniques to generate more accurate recommendations. This helps to provide customers with a more tailored user experience as they are presented with more relevant product recommendations.

*d) Increased Product Awareness:* By providing customers with more tailored product recommendations, it can help to increase product awareness and promote products to customers who may not have otherwise been aware of them. This can help to increase sales and revenues for the company.

2) *Diadvantages:*

*a) Cost:* Hybrid e-commerce recommender systems can be expensive to develop and maintain since they require to be efficient, many technologies must be combined, including machine learning, natural language processing, and data mining.. Additionally, depending on the size of the system, cost of hosting and support services may be required.

*b) Complexity:* Building a hybrid e-commerce recommender system requires a lot of technical expertise, which can be difficult to acquire. Additionally, the system can be complex to maintain and troubleshoot, and requires ongoing optimization.

*c) Data Quality:* The quality of the data used to build and maintain the system is critical to its success. Poor quality data can lead to inaccurate recommendations, which can reduce customer satisfaction and ultimately hurt the business.

*d) Privacy:* As with any system that collects and stores customer data, privacy is a major concern. It is important to ensure that customer data is handled securely and in accordance with applicable laws and regulations.

V. THE CHALLENGES FOR BIG DATA-BASED E-COMMERCE RECOMMENDER SYSTEM

E-commerce recommender systems are becoming increasingly important in helping customers find the product that best suits their needs or desires. With the rise of big data, e-commerce recommender systems have become increasingly sophisticated, allowing for personalized product recommendations tailored to individual customers' preferences or interests.

In data-intensive systems, big data calls for robust data processing procedures. Velocity, Volume, Variety, and Veracity, or 4V qualities, are often used to describe big data. These characteristics cover data processing speed, data storage capacity, data structure (structured, semi-structured, or unstructured), provenance, and curation. This technique alludes to the Hadoop data-scale framework. Every second, data sets for e-commerce are created that encompass all different kinds of data, including data on items, usage, transactions, comments, ratings, and real-time data. The following issues for e-commerce recommender systems are brought on by big data.

A. *Data Privacy:* Storing and using personal data for recommendation purposes can raise privacy and security concerns related to the use and misuse of personal data.
B. *Cold Start Problem:* It is difficult to recommend products to new users who have not interacted with the system before, as there is no data to draw on.
C. *Quality of Recommendations:* The effectiveness of the suggested outcomes is dependent on the data that is used and the algorithms that are employed.
D. *Scalability:* The system need to be capable of handling a lot of data and provide accurate recommendations in real-time.
E. *Robustness:* The system should be robust enough to handle distinct user categories and different data kinds.
F. *Personalization:* The system should be able to personalize recommendations based on individual user preferences.

VI. WAYS TO OVERCOME THE EXISTING CHALLENGES OF THE RECOMMENDER SYSTEM BASED ON BIG DATA

The use of Big Data in Recommender Systems has increased significantly in recent years. Big Data has enabled companies to make better decisions and better serve their customers. However, there are still the difficulties of using big data in Recommender Systems. These challenges include data quality issues, privacy concerns, scalability issues, and algorithmic complexity.

A. *Limited Resources: -*

The limited resources problem is a challenge that arises when trying to create a recommender system. This problem occurs when there is not enough information available about users or items to make accurate recommendations. This can be a result of data sparsity, which is a common issue in recommender systems. To combat the limited resources problem, recommender systems often rely on user feedback to help improve recommendations.

1) *There are a few ways to overcome the issue of limited resources in a e-commerce recommender system:*

*a)* Use a hybrid recommender system that combines different types of recommender systems. For example, a content-based recommender system can be used depending on their previous purchasing behaviour, to suggest products to consumers.

*b)* Use a user-based recommender system that makes product recommendations to users based on the items that other users who share similar interests have bought.

*c)* Use a recommender system that bases its recommendations on the goods that users have previously purchased.

B. *Data Validity Period : -*

There are a few different types of data validity problems that can occur in recommender systems. One type of problem is when the data that is used to train the system is no longer valid when the system is actually used. This can happen when the data used to train the system is collected over a period of time, but the system is used over a different period of time. For example, if a recommender system is trained on data from January to June, but it is used from July to December, the data may no longer be valid. Another type of data validity problem can occur when the data used to train the system is not representative of the data that will be used when the system is actually used. This can happen when the training data is gathered from an alternative source population than the population that will be using the system. For example, if the training information is gathered from people who live in a certain city, but the system will be used by people who live in a different city, the data may not be representative. yet another type of data validity problem can occur when the data used to train the system is biased. This can happen when the data is collected in a way that is not random. For example, if the data is collected from people who are more likely to use the system, the data will be biased.

*1) There are a few ways to overcome the data validity period challenge in recommender system for e-commerce:*

*a)* Use a data validation technique such as cross-validation to ensure that the model is not overfitting to the data.

*b)* Use a data pre-processing technique such as feature scaling to ensure that the data is standardized and can be easily compared.

*c)* Use a data augmentation technique such as adding synthetic data to the training set to increase the amount of data available for training.

*d)* Use a data selection technique such as selecting only the most recent data for training to ensure that the model is trained on the most relevant data.

*C. Cold Start: -*

The cold start problem when a recommender system fails to deliver recommendations for a new user or new item due to a lack of sufficient data to do so. This can be a major issue for recommender systems because their primary purpose is to make recommendations based on past user behavior. By utilising a hybrid recommendation engine that incorporates content-based and collaborative filtering techniques, the cold start issue can be solved.

*1) There are a few ways to avoid a cold start challenge in recommender systems for e-commerce:*

*a) Use content-based filtering:* This approach relies on the user's past behavior (e.g. what items they have viewed or purchased) to make recommendations. This is effective for overcoming the cold start problem because it does not require other users' data in order to make recommendations.

*b) Use a hybrid approach:* This approach combines content-based filtering with collaborative filtering. This can be done by using content-based filtering to make initial recommendations, and then using collaborative filtering to fine-tune the recommendations.

*c) Use social media data:* If a user is connected with their friends on social media, their friends' data can be used to make recommendations. This is effective for overcoming the cold start problem because it allows recommendations to be made even if the user does not have any past behavior data.

*d) Use demographic information:* The user's age, gender, location, and other details are taken into account in this approach while making recommendations. This is effective for overcoming the cold start problem because it does not require other users' data in order to make recommendations.

*D. Long Tail Problem: -*

A long tail recommender system is a specific kind of recommender system that can recognise and recommend items from a long tail of items. This is in contrast to traditional recommender systems, which are typically only able to recommend items from a limited set of items.
Recommendation algorithms in long tail problem refers to the challenge of making accurate suggestions for products that are not well-known or well-liked. This is true because traditional recommender systems rely on collaborative filtering, which is predicated on the idea that people with similar likes will appreciate things that are similar to them. However, this assumption does not hold for long tail items, which are often obscure and not well-known. As a result, traditional recommender systems often fail to provide accurate recommendations for long tail items.
Finally, it is also possible to use a long tail recommender system, which is specifically designed to identify and recommend long tail items. These systems typically use a combination of content-based filtering and hybrid methods.

*1) There are a number of ways to overcome the long tail problem in recommender systems:*

*a)* Use a a hybrid strategy that incorporates content-based filtering with collaborative filtering. This way, you can recommend items to users even if there is limited data on them.

*b)* Use a technique called latent factor analysis to find hidden patterns in the data. This can help you make better recommendations to users even if there is limited data on them.

*c)* Use a technique called item-based collaborative filtering. This approach looks at items that are similar to each other and makes recommendations based on that. 4. Use a technique called user-based collaborative filtering. With this approach, recommendations are based on users who are similar to one another..

*E. Scalability: -*

A scalability issue in recommender system is when the system is not able to handle an increased number of users or items. This can happen when the system is not designed to handle large numbers of users or items, or when the system slows down as the number of users or items increases. This can be a problem for users who are trying to use the system with a large number of friends or items, or for developers who are trying to add new users or items to the system.

*1) There are a few ways to overcome the scalability issue in Recommender system:*

   *a) Use a scalable algorithm:* A scalable algorithm is one that can handle large datasets without slowing down. There are a few scalable algorithms available, such as Apache Hadoop and Spark.

   *b) Use a distributed system:* A distributed system is a system that is spread across multiple machines. This can help with scalability as each machine can handle a portion of the data.

   *c) Use a cloud-based system:* A cloud-based system is one that is hosted on a remote server. This can be scalable as the remote server can have a lot of resources.

## VII. RECOMMENDER SYSTEM BASED ON E-COMMERCE EXAMPLES

### A. Amazon.com

The algorithm Amazon uses is the combination of two approaches that is Collaborative filtering and Content-Based filtering known as Item-Based Collaborative Filtering. Amazon creates a unique online store just like in a walk-in store the clerk would recommends the product for each customer viewing different product based on their interests. Amazon also stores the searches of other customers that purchases the same product. Every recommendations are based on a variety of variables, including location, previous purchases, trends, saved goods, and more deals or discounts, purchases made by other customers after browsing comparable products, user reviews, and curated top selections for customers[13].

### B. Drugstore.com

The Adviser feature at Drugstore.com allows the customer to purchase a product from any category such as skincare, common flu and so on to indicate their preferences. For example, in the high fever, the customers show the symptoms such as headache and aching muscles, what type of medications they need and the age of the patient. With the provided information the Advisor recommends the product list to meet the conditions.

### C. MovieFinder.com

Like many online websites MovieFinder.com allows the customer to register with the website. The customer receives editors' recommendations in a category of their choice. Customers are required to choose a category from a list of pre-existing categories. After choosing a list, the buyer is guided by an editor's summary of the best films.

### D. eBay

Both buyers and sellers can submit customer feedback profiles on eBay.com for other customers they have transacted with. Ratings and comments are both forms of feedback. Buyers, who can read the profile of sellers, are given a recommender system using feedback. Customers can also search the seller-specific ratings and reviews upon request. Customers can indicate the products they are interested in buying using eBay's personal shopper option. Customers enter a "short term" (i.e., 30/60/90 days) and perform a search using a list of keywords, including a price limit, of their choice. The website conducts the client's search over all of the site's auctions on a periodic basis (every one or three days), and then sends the consumer an email with the results.

## VIII. CONCLUSION AND FUTURE SCOPE: -

In this paper, we covered a number of widely utilized E-commerce recommendation techniques, including collaborative filtering, content-based filtering, and recommendation systems based on community and demographic data. Through this paper, we have concluded that the recommendation systems have evolved into a important tool of many e-commerce websites such as Amazon, Netflix and so on. The recommendation applications such as recommending movies, grocery, medicines and so on. Instead of purchasing the product from the store physically now people prefer online websites. For example, reading a book from millions of books available on the web instead of choosing a book from a store physically. Recommendation systems has made the task easier for the customers by listing the products of their interest. It has turned the visitor's websites into actual buyers. But still recommender systems face challenges such as cold start, scalability, limited resources, long tail problems and data validity period. There are numerous approaches to enhance recommender systems for e-commerce. One way is to make the systems more personalised, so that they take into account the individual preferences of users. Another way is to make the systems more contextual, so that they take into account the user's current situation and how the purchase was made in context. Finally, Recommendation engines made more effective by integrating them with other systems, such as search engines, to provide a more comprehensive service.